\documentclass[twoside,twocolumn,10pt]{article}

\usepackage{amsfonts}
\usepackage{amsmath}

\usepackage{wscg}           
\RequirePackage{ifpdf}
\ifpdf
 \RequirePackage[pdftex]{graphicx}
 \RequirePackage[pdftex]{color}
\else
 \RequirePackage[dvips,draft]{graphicx}
 \RequirePackage[dvips]{color}
\fi

\usepackage{nopageno}       

\usepackage{amsthm,amssymb}

\title{$\Pi$-surfaces: products of implicit surfaces towards constructive composition of 3D objects}

\author{
\parbox{1.0\textwidth}{\centering
Adriano N. Raposo and Abel J.P. Gomes\\[1mm]
Instituto de Telecomunica\c{c}\~{o}es and
Universidade da Beira Interior\\
R. Marqu\^{e}s de \'{A}vila e Bolama\\
6200-001, Covilh\~{a}, Portugal\\[1mm]
anraposo@ubi.pt, agomes@di.ubi.pt
}
}

\usepackage{url}
\makeatletter
\def\Uslash{\mathbin{\mathchar`\/}\@ifnextchar{/}{\kern-.15em}{}}
\g@addto@macro\UrlSpecials{\do \/ {\Uslash}}
\def\Ucolon{\mathbin{\mathchar`:}\@ifnextchar{/}{\kern-.1em}{}}
\g@addto@macro\UrlSpecials{\do : {\Ucolon}}
\makeatother

\begin{document}

\twocolumn[{\csname @twocolumnfalse\endcsname

\maketitle  

\begin{abstract}
\noindent

Implicit functions provide a fundamental basis to model 3D objects, no matter they are rigid or deformable, in computer graphics and geometric modeling. This paper introduces a new constructive scheme of implicitly-defined 3D objects based on products of implicit functions. This scheme is in contrast with popular approaches like  \emph{blobbies}, \emph{meta balls} and \emph{soft objects}, which rely on the sum of specific implicit functions to fit a 3D object to a set of spheres.

\end{abstract}

\subsection*{Keywords}
Implicit surfaces, surface modeling, constructive modeling, 3D modeling, computational geometry, geometric modeling.

\vspace*{1.0\baselineskip}
}]


\section{Introduction}
\label{sec:introduction}

\copyrightspace

Both implicit surfaces and parametric surfaces have been widely used in computer-aided design, geometric modeling, visualization, animation, and computer graphics \cite{Gomes09}. Implicit surfaces benefit from the valuable properties in modeling, namely: closure and point membership. Indeed, applying boolean operations (intersection, union, and difference) results in another well-defined implicit surface \cite{1999-gomes,Barbier04}. Moreover, it is easy to check whether a point behind, on, or beyond an algebraic implicit surface, as needed in collision detection. 

However, unlike piecewise parametric surfaces, implicit surfaces are not akin to local shape changes interactively. Also, rendering an implicitly-defined algebraic surface poses some difficulties because it requires its preliminary triangulation \cite{Requicha85,Pilz89,2003-gomes}, particularly for implicit surfaces defined by high degree polynomial functions; hence, the low degree algebraic surface patches or algebraic splines used for geometric modeling \cite{babaj88,babaj97,lazard,chen, warren, wu}.

In this paper, we introduce a new method to model complex shapes through products of implicitly-defined algebraic surfaces (e.g., spheres, ellipsoids, cylinders, hyperboloids, and tori), called $\Pi$-\emph{surfaces}. These $\Pi$-surfaces allow us: (1) to represent a surface as the product of two or more patches; (2) to globally edit the shape of each patch (e.g. patch replacement, patch removal, patch insertion);
(3) to locally edit the shape of each patch (e.g. deformations like recesses, saliences, and so forth); (4) to blend two patches using a single blending parameter; and (5) to better control bulging effects inherent to sum-based surfaces (e.g., $\sum$-surfaces like Gaussian surfaces) via the blending parameter.

In short, we propose a new constructive method for 3D objects through products of implicit functions, which differs from the dominant method found in the literature, which is based on $\sum$-surfaces (e.g., \emph{blobby molecules} \cite{blinn82}, \emph{metaballs} \cite{Nishimura85}, \emph{soft objects} \cite{Wyvill86}, \emph{blobby model} \cite{Muraki91}, piecewise implicit surface patches  \cite{babaj88,babaj97}, and homotopy-generated surfaces \cite{Bedi92} \cite{Hoffmann85}.

\section{$\Pi$-surfaces}
\label{sec:method}

\emph{$\Pi$-surfaces} can be used to approximate any given smooth and bounded object in $\mathbb{R}^3$ whose surface is defined by a single polynomial as a product of subsidiary polynomials. In other words, we can design any smooth object with a single algebraic surface. Let us denote the defining polynomials as $f_i\in\mathbb{R}[x_1,\ldots,x_n](i=1,\ldots,k)$. Then, the approximating object is defined by the polynomial

\begin{equation}
\label{eq:pi-surface}
F(x,y,z) = \prod_i f_i(x,y,z) - r
\end{equation}

\noindent where $r\in\mathbb{R}$ stands for the blending parameter that controls the approximating error.

\begin{figure*}[ht]
\begin{center}
\includegraphics[width=\textwidth]{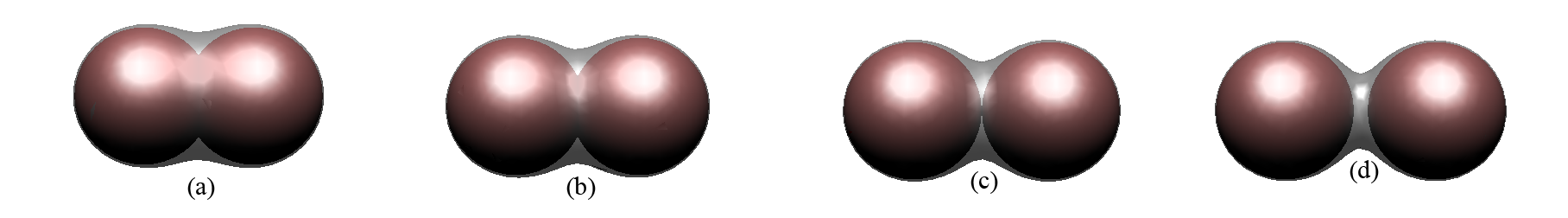}
\caption{\small{Repositioning of two spheres of the surface $F(x,y,z)=((x-d)^2+y^2+z^2-1)\cdot((x+d)^2+y^2+z^2-1)-0.25$: (a) $d=0.8$; (b) $d=0.9$; (c) $d=1.0$ and (d) $d=1.1$. The positions of the surface primitives were adjusted, but the parameter $r$ remained unchanged. The $\Pi$-surface appears in translucent gray.}} \label{fig:two_balls}
\end{center}
\end{figure*}

\begin{figure*}[t]
 \begin{center}
 \includegraphics[width=.75\textwidth]{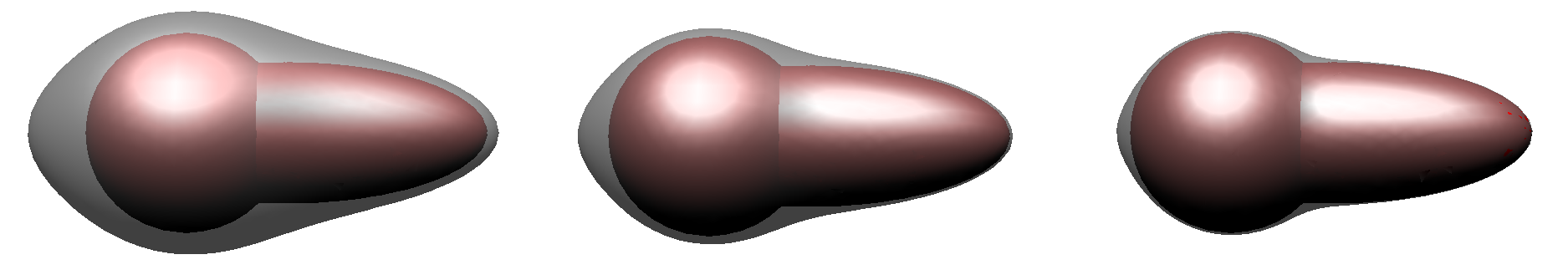}\\
 (a) ~~~~~~~~~~~~~~~~~~~~~~~~~~~~~~~~~~~~~~~~~ (b) ~~~~~~~~~~~~~~~~~~~~~~~~~~~~~~~~~~~~~~~~~ (c)
 \caption{\small{``Eggplant'' surface defined  by the function  $F(x,y,z)=(\frac{1}{64}\,x^2+\frac{1}{8}\,y^2+\frac{1}{8}\,z^2-0.25)\cdot((x+2)^2+y^2+z^2-4)-r=0$ for different values of $r$: (a) 1.0; (b) 0.25; and (c) 0.05. The $\Pi$-surface appears in translucent gray.}} \label{fig:eggplant3}
\end{center}
\end{figure*}

\begin{figure*}[ht]
\begin{center}
\includegraphics[width=.75\textwidth]{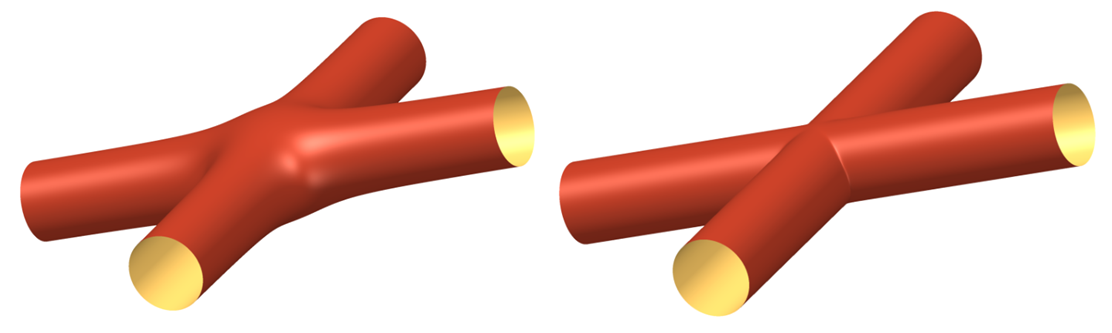}\\
(a) ~~~~~~~~~~~~~~~~~~~~~~~~~~~~~~~~~~~~~~~~~~~~~~~~~~~~~~~~~~~~ (b)
\caption{\small{The parameter $r$ is used to reduce an undesired bulge that occurs about the intersection of two crossed cylinders defined by the function  $F(x,y,z)=(x^2+z^2-1)\cdot(y^2+z^2-1)-r=0$: (a) $r=1$; (b) $r=0.001$}} \label{fig:bulges}
\end{center}
\end{figure*}

From Eq.~(\ref{eq:pi-surface}), which is the core equation of this paper, we can find that the shape of the approximating surface depends on the primitive surfaces $f_i$ and the parameter $r$. 
That is, we have a product ($\Pi$) of functions $f_i$, where each function represents an arbitrary geometric primitive $i$, and $r$ is the approximation parameter for the entire surface.

\subsection{Implicit Primitives}
\label{sec:primitives}

\emph{$\Pi$-surfaces} are built up from arbitrary implicit surface primitives. However,  unlike traditional distance-based methods, $\Pi$-surface primitives are not limited to sets of spheres (or ellipsoids) to compose 3D objects. Instead, $\Pi$-surfaces use primitives like planes, quadrics, tori, and other more complicated surfaces. These primitives split into two different groups: \emph{bounded primitives} (e.g.,  sphere, ellipsoid, torus, and so forth) and \emph{unbounded primitives} (e.g., plane, cylinder, cone, hyperboloid of 1 sheet,  hyperboloid of 2 sheets, paraboloid, hyperbolic paraboloid, and the like). Both bounded and unbounded primitives may be combined into the same 3D object.

\subsection{Shape Operations}
\label{sec:deformation}

There are three different types of shape operations: \emph{shape repositioning}, \emph{global shrinking or inflation}, and \emph{local bulges or concavities}.

\subsubsection{Shape Repositioning}
\label{subsec:repos}

The shape of a $\Pi$-surface can also be adjusted by controlling the position of the primitives. This is somehow intuitive, though the blending of the primitives only depends on the distance between them and the blending parameter $r$. For example, both spherical primitives in Figure~\ref{fig:two_balls} are combined into a $\Pi$-surface (in transparent gray). Obviously, considering the blending parameter $r=0.25$ remains unchanged, the resulting $\Pi$-surface depends on the distance between both primitives.

\subsubsection{Global Shrinking vs. Global Inflation}
\label{subsec:r}

Varying the blending parameter $r$ in Eq.~(\ref{eq:pi-surface}) makes the entire $\Pi$-surface shrink or inflate. Specifically, decreasing the value of $r$ to zero makes the $\Pi$-surface shrink to the surface of the union of implicit surface primitives. Conversely, increasing the value of $r$ inflates the $\Pi$-surface globally. Figure~\ref{fig:eggplant3} shows how distinct values of $r$ affect the global shape of a $\Pi$-surface.

\begin{figure*}[ht]
\begin{center}
\includegraphics[width=.75\textwidth]{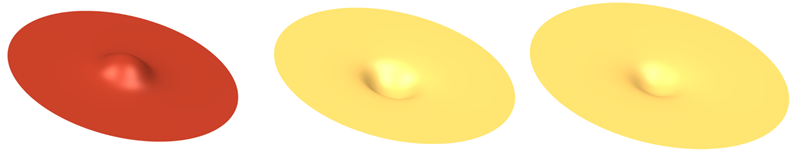}\\
\quad (a) ~~~~~~~~~~~~~~~~~~~~~~~~~~~~~~~~~~~~~~~~~ (b) ~~~~~~~~~~~~~~~~~~~~~~~~~~~~~~~~~~~~~~~~~ (c) 
\caption{\small{Local deformations obtained by placing a small sphere near the $x=0$ plane : (a) protrusion defined by a sphere in $(f(x,y,z)=(x^2+y^2+z^2-1)\cdot(x)-1=0$; (b) depression obtained by inverting the sign of the original plane as in  $f(x,y,z)=(x^2+y^2+z^2-1)\cdot(-x)-1=0$; and (c) concavity obtained by inverting the sign of the sphere as in $f(x,y,z)=(-x^2-y^2-z^2-0.01)\cdot(x)-1=0$ (right)}} \label{fig:concavities}
\end{center}
\end{figure*}

\begin{figure*}[ht]
\begin{center}
\includegraphics[width=.75\textwidth]{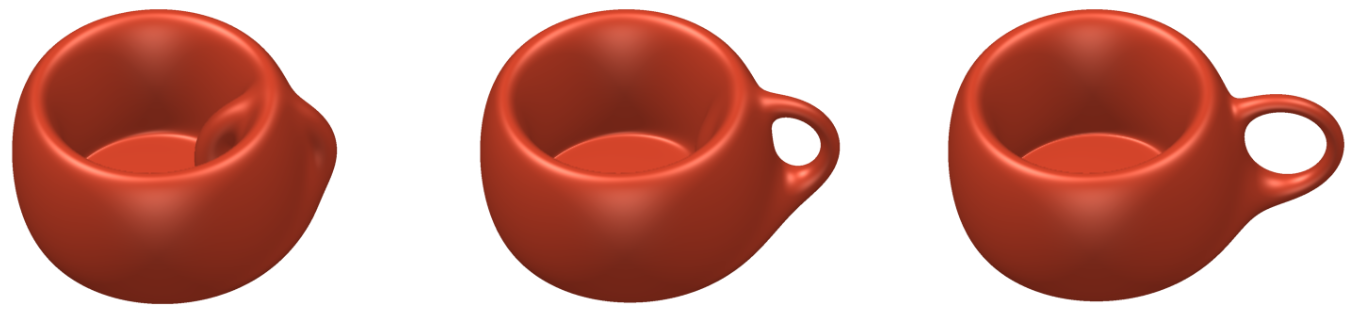}\\
(a) ~~~~~~~~~~~~~~~~~~~~~~~~~~~~~~~~~~~~~~~~~ (b) ~~~~~~~~~~~~~~~~~~~~~~~~~~~~~~~~~~~~~~~~~ (c) ~~~~~~~~
\caption{Interactive edition of a moving handle (or torus) to create a cup as given by the function $F(x,y,z)=(((x-2.2 \cdot a)^2+(z+1)^2+y^2+0.5-0.01)^2-2 \cdot ((x-2.2 \cdot a)^2+(z+1)^2)) \cdot ( \frac{1}{8} \cdot x^2+ \frac{1}{8} \cdot y^2+(z+2)^2-0.1) \cdot (((z+1)^2+10 \cdot x^2+10 \cdot y^2+19)^2-800 \cdot (x^2+y^2))-10$.}
\label{fig:interactive_cup}
\end{center}
\end{figure*}

Similar to $\sum$-surfaces, undesired bulges about the intersection regions of  primitives may occur in $\Pi$-surfaces. The difference is that $\Pi$-surfaces allow the control on bulges by reducing the value of the approximation parameter $r$, as illustrated in Figure~\ref{fig:bulges}.

\subsubsection{Local Deformations}
\label{subsec:concavities}

Unlike $\sum$-surfaces, $\Pi$-surfaces are very efficient in performing local deformations without affecting the global shape of the objects. Such deformations are performed by the action of primitives on other primitives.
There are two main types of local deformations: \emph{local protrusions}  and \emph{local depressions}.

\emph{Local protrusions} are generated by adding a relatively small primitive to the $\Pi$-surface, but the small primitive needs to be close enough to the $\Pi$-surface in order to affect it. This mechanism is illustrated by the red object in  Figure \ref{fig:concavities}(a), where a small sphere imposes a spherical bump to the plane $x=0$.

\emph{Local depressions} are a little bit trickier to obtain with $\Pi$-surfaces. In this case, we need to invert the sign of the implicit function of the small primitive, yet inverting the original implicit function of the global shape is also feasible. This local deformation mechanism is illustrated in Figure \ref{fig:concavities}(b)-(c), where we also used a small sphere to obtain a concavity in the plane $x=0$.

\subsection{Interactive Shape Control}
\label{sec:interactive_shape_control}

As seen above, adding a primitive is as simple as including its defining implicit function in the product of $\Pi$-surface. This procedure only changes the already existing object locally. Besides, the position and geometric parameters of the primitive can also be interactively adjusted on the fly by the user. There two types of shape composition control: additive and subtractive.  

\emph{Additive shape composition} means we interactively add positive primitives to a given object without losing the control of its overall shape, i.e., without losing the capability of anticipating how the inclusion of a specific primitive will affect the global shape the object, as illustrated in  Figure~\ref{fig:interactive_cup}. 
In this case, the user creates the handle of the cup by interactively adding a torus to an existing bowl composed by a flattened ellipsoid and another torus. As depicted in Figure~\ref{fig:interactive_cup}, when the user interactively displaces the handle until its final position, the handle remains perfectly blended with the cup. If needed, the user might also interactively change the ''thickness'' of the handle just by changing the smaller radius of the torus.

\begin{figure*}[ht]
\begin{center}
\includegraphics[width=.20\textwidth]{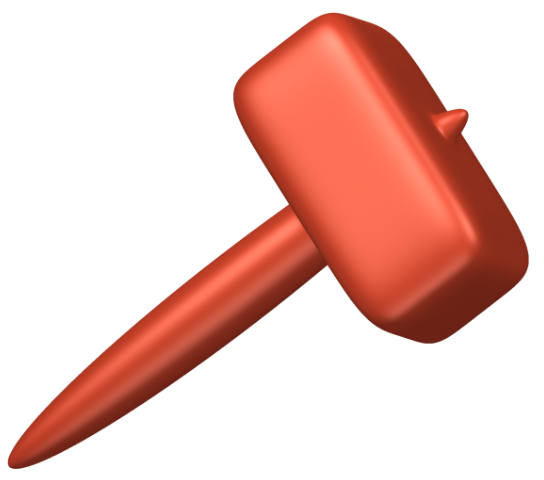}
\includegraphics[width=.34\textwidth]{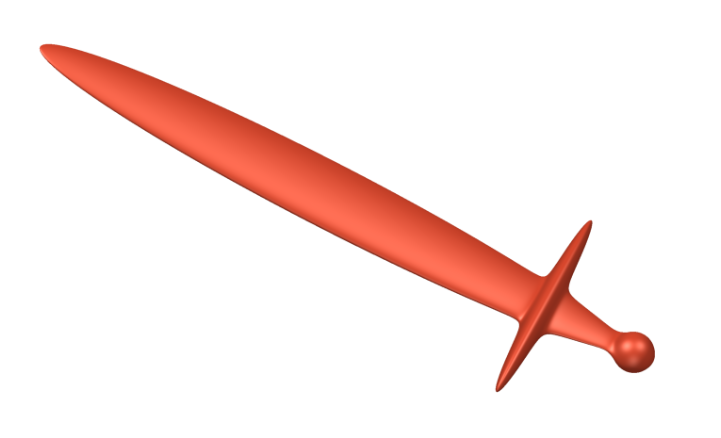}
\includegraphics[width=.40\textwidth]{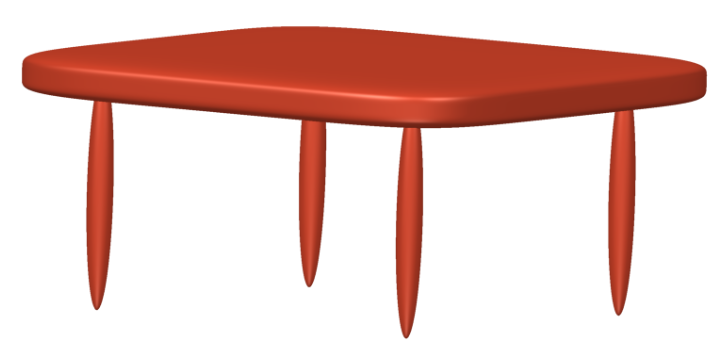} \\ 
(a) \quad\quad\quad\quad\quad\quad (b) \quad\quad\quad\quad\quad\quad (c)
\caption{\small{Additive shape composition of $\Pi$-surface objects: (a) hammer; (b) sword; and (c) table.}} \label{fig:examples}
\end{center}
\end{figure*}

\begin{figure*}[ht]
\begin{center}
\includegraphics[width=.75\textwidth]{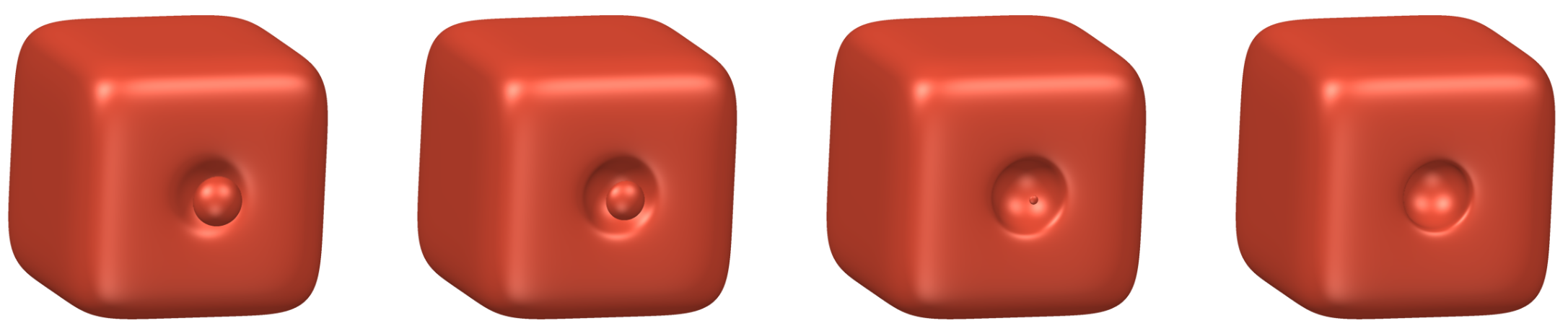}
\caption{Interactive edition of a depression on a rounded cube surface by moving a small negative sphere. The resulting object is defined by the following $\Pi$-surface: $F(x,y,z)=(x^6+y^6+z^6-1) \cdot (-(x+a)^2-(y+b)^2-(z+c)^2+0.05)-0.01$, where $(a,b,c)$ is the position of the sphere. }
\label{fig:interactive_cube}
\end{center}
\end{figure*}

Let us now see how three more objects are obtained from the composition of positive primitives.

\emph{Example 1}. The hammer shown in Figure \ref{fig:examples}(a) was built using two  implicit primitives so that $F=f_1 \, . \, f_2 - c$, where $c=1$ is the blending parameter, and $f_1$ defines the first primitive (a block) and $f_2$ the second primitive (an ellipsoidal handle) as follows:
$$f_1 = (x-5)^6 + \left( \frac{y}{2} \right) ^6 + z^6 - 100; f_2 = \left( \frac{x}{8} \right) ^2 + y^2 + z^2 - 1.$$

\emph{Example 2}. The sword pictured in Figure \ref{fig:examples}(b) is defined by $F=f_1 \, . \, f_2 \, . \, f_3 - 5$, where $f_1$, $f_2$, and $f_3$ define three implicit primitives,  two crossed ellipsoids and a sphere, as follows:
$$ f_1 = (x+17)^2+y^2+z^2-1; $$
$$ f_2 = (2x+24)^2 + \left( \frac{y}{6} \right) ^2+z^2-1; $$
$$ f_3 = \left( \frac{x}{16} \right) ^2 + \left( \frac{y}{2} \right) ^2 + (2z)^2-1. $$

\emph{Example 3}. The table shown in Figure \ref{fig:examples}(c) is defined by 
$F = f_1 \cdot f_2 \cdot f_3 \cdot f_4 \cdot f_5 - 1$, that is, by five  implicit primitives, 
a rounded parallelepiped for the table top and four ellipsoids for the table legs, where
$$ f_1 = 0.1 \cdot x^6 + 0.1 \cdot y^6 + (8z+5)^6 - 1; $$
$$ f_{2,3,4,5} = \left( \frac{x \pm 1}{0.1} \right) ^2 + \left( \frac{y \pm 1}{0.1} \right) ^2 + \left( \frac{z}{0.9} \right) ^2 - 0.5. $$

\emph{Subtractive shape composition} means we interactively add negative primitives to an existing $\Pi$-surface, as shown in Figure~\ref{fig:interactive_cube}.
Interestingly, the negative primitive used to create the concavity shrinks as it gets close to the existing object, and even vanishes at some point.
This is a very useful feature because it eliminates undesired components in the final object; specifically, the negative sphere in Figure \ref{fig:interactive_cube} is used to interactively create a concavity in one of the faces of a 3D cube. Notice how the sphere becomes smaller and smaller to a point at which it finally vanishes in the proximity of the cube.

\subsection{Geometry Evaluation}

The geometry evaluation (also known as boundary evaluation \cite{Requicha85}) builds upon ray casting to solve two difficult problems in implicit surface modeling: 
\begin{itemize}
\item \textit{Rendering}.
We can immediately visualize the surface on screen with a significant realism and geometry fidelity, including singularities like apices, intersection curves, and the like, though without explicitly solving such singularities (i.e. the loci where the first derivatives vanish), as well as multi-component surfaces (see, for example, \cite{Raposo06}).
\item \textit{Triangulation}.
The point sampling inherent to ray casting allows us to reconstruct the surface  because we classify the surface points as either regular or singular using the gradient (i.e. first derivatives). Essentially, this problem consists in triangulating a point cloud whose points are samples of the surface (see, for example, \cite{Bernardini99,Levet05,Kazhdan06}). Note that the nearby points are associated with adjacent pixels on the screen so that the triangulation is more amenable to carry out than in standard surface reconstruction algorithms lacking information like normals or partial derivatives. Moreover, knowing the partial derivatives at such points, we can quickly identify singularities like apices, sharp features, self-intersections, and so forth. Summing up, we use a novel image-based surface reconstruction, though we also may use algorithms like those described by Skala et al. in \cite{Skala04,Skala05,Skala07}. Obviously, after triangulating a surface, we no longer need ray casting for the purpose of the graphics output.
\end{itemize}

Thus, our renderer combines ray casting and triangulation approaches into a single renderer.  

\subsection{Molecular $\Pi$-surfaces}
\label{sec:molecular}

Molecular modelling is an important application for  $\Pi$-surfaces. Knowing that each atom has a corresponding \emph{van der Waals} (VDW) radius \cite{bondi64}\cite{batsanov01}, a molecule can be seen as a set of overlapping spheres representing the atoms. 

\subsubsection{Molecular $\Pi$-surface Formulation}

The molecular surface $S$ of a molecule can be then described by the following $\Pi$-surface:

\begin{equation}
\label{eq:molecular_surface}
S = \prod_{i=1}^{n}f_i-r 
\end{equation}
where 
\begin{equation}
\label{eq:atomicfunction}
f_i= (x-x_i)^2+(y-y_i)^2+(z-z_i)^2 - {R_i}^2
\end{equation}
represents the spherical surface of the $i$-th atom of radius $R_i$ and center $(x_i,y_i,z_i)$. This formulation is in contrast with the Gaussian $\Sigma$-surface proposed by Blinn \cite{blinn82} to represent molecular surfaces. 

It is worthy noting that for $r>0$, Equation \ref{eq:molecular_surface} represents a smooth molecular surface, so there is no need to use ray casting to sample the surface. Instead, we can directly triangulate the surface on-the-fly using triangulation algorithms as those described in  \cite{Raposo06,raposo09,dias11,dias15,dias17}.

\subsubsection{Examples of Molecular $\Pi$-surfaces}

Let us now see how to model the molecular structures of two real drugs using $\Pi$-surfaces: (a) \emph{formic acid} and (b) \emph{acetic acid}. 

\emph{Formic acid}. This drug occurs naturally and is present in the venom of bees and ant stings. Using the atomic structure obtained from the DrugBank and the VDW radii from \cite{batsanov01} (1.52 \AA~for oxygen and 1.70 \AA~for carbon), the spheres have the following center points and radii:

$$ c_1 = (2.310,-1.334,0.000), R_1 = 1.52 ; $$ 
$$ c_2 = (3.644,-2.104,0.000), R_2 = 1.70 ; $$ 
$$ c_3 = (4.977,-1.334,0.000), R_3 = 1.52. $$

Then, using Equation \ref{eq:molecular_surface} (with $r=1$) we obtain the $\Pi$-surface that represents the \emph{formic acid} molecule as follows:

\begin{align*} 
 S=((x-2.310)^2+(y+1.334)^2+z^2-1.52^2) \cdot \\((x-3.644)^2+(y+2.104)^2+z^2-1.70^2) \cdot \\((x-4.977)^2+(y+1.334)^2+z^2-1.52^2) - 1
\end{align*}

\begin{figure}[ht]
\begin{center}
\includegraphics[width=.5\textwidth]{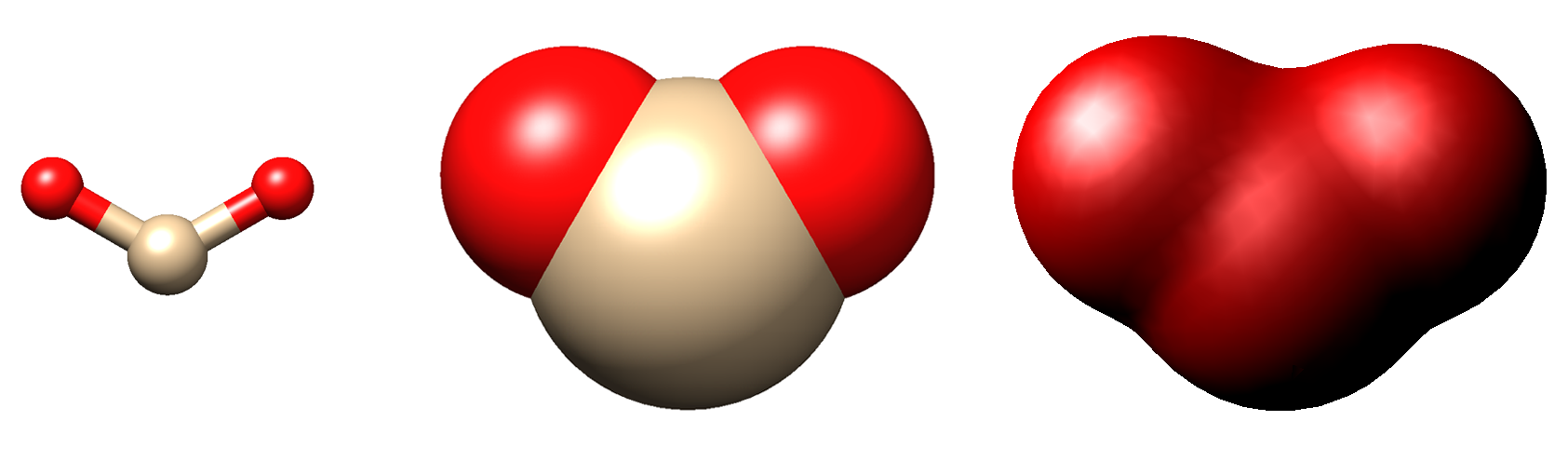}\\
(a) ~~~~~~~~~~~~~~~~~~~~~~~~
(b) ~~~~~~~~~~~~~~~~~~~~~~~~
(c) ~~~\\
\caption{\small{Formic acid: (a) ball-and-stick; (b) atom-describing  spheres (VDW representation); and (c) molecular $\Pi$-surface.}} \label{fig:formic_acid}
\end{center}
\end{figure}

The formic acid $\Pi$-surface is shown in Figure \ref{fig:formic_acid} (c).

\emph{Acetic acid}. This drug results from the oxidation of ethanol and destructive distillation of wood. As seen in the previous example, the center points and radii of its atoms are as follows:

$$ c_1 = (24.214,-24.150,0.000), R_1 = 1.70 ; $$
$$ c_2 = (25.548,-23.380,0.000), R_2 = 1.70 ; $$
$$ c_3 = (26.881,-24.150,0.000), R_3 = 1.52 ; $$
$$ c_4 = (25.548,-21.840,0.000), R_4 = 1.52 $$

Then, using Equation \ref{eq:molecular_surface} (with $r=1$), we obtain the algebraic surface that represents the \emph{acetic acid} molecule as follows:

\begin{align*} 
S=((x-24.214)^2+(y+24.150)^2+z^2-1.70^2) \cdot \\((x-25.548)^2+(y+23.380)^2+z^2-1.70^2) \cdot \\((x-26.881)^2+(y+24.150)^2+z^2-1.52^2) \cdot \\((x-25.548)^2+(y+21.840)^2+z^2-1.52^2) - 1
\end{align*}

\begin{figure}[ht]
\begin{center}
\includegraphics[width=.5\textwidth]{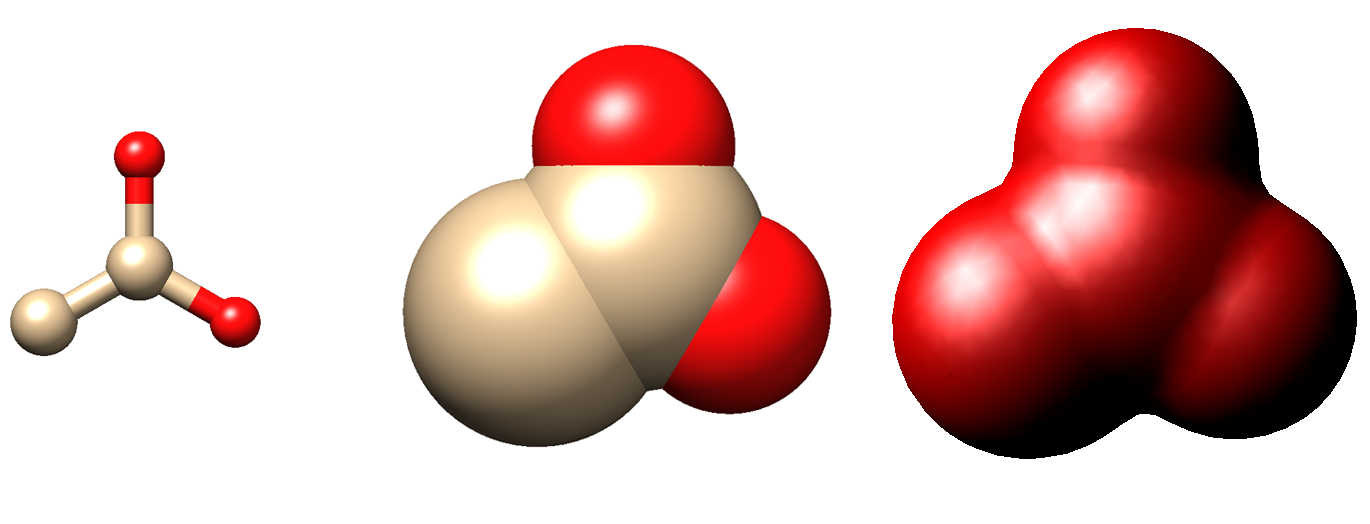}\\
(a) ~~~~~~~~~~~~~~~~~~~~~~~~~~~
(b) ~~~~~~~~~~~~~~~~~~~~~~~~~~~
(c) ~~~\\
\caption{\small{Acetic acid: (a) ball-and-stick; (b) atom-describing spheres (VDW representation); and (c) molecular $\Pi$-surface.}} \label{fig:acetic_acid}
\end{center}
\end{figure}

The \emph{acetic acid} $\Pi$-surface is depicted in Figure \ref{fig:acetic_acid} (c).

The \emph{Molecular $\Pi$-surface} formulation can also be used, for the sake of simplification, to model the four nucleotides (adenine, cytosine, guanine and thymine) used as DNA building blocks in \cite{Raposo12,Raposo14,Raposo15}, replacing the Gaussian $\Sigma$-surface model.

\section{Conclusions and Future Work}
\label{sec:conclusions}

This paper presents a new general method to model complex 3D objects as the product of algebraic functions representing simpler implicit surface primitives. The resulting surfaces are here called $\Pi$-surfaces. Unlike traditional methods used in implicit modeling, this new method is not based on a sum of specific distance functions. Instead, the 3D objects are built upon a product of algebraic functions representing arbitrary implicit surfaces such as spheres, ellipsoids, and  tori. This new approach allows us to interactively perform controlled deformations on the object shape both locally and globally. Two possible applications are proposed for this new model: \emph{constructive building of 3D objects} and \emph{molecular surfaces modeling}. As future work, and knowing that an increase in the number of primitives might generate higher degree algebraic surfaces, it is our intention to develop some strategies to overcome possible limitations that may occur in these cases. 
As a final remark, we believe that this new method is an important contribution for the geometric modeling research field.  

\section{Acknowledgements}

This research has been partially supported by the Portuguese Research Council (Funda\c c\~ao para a Ci\^encia e Tecnologia), under the FCT Project UID/EEA/50008/2019. We are also grateful to reviewers for their suggestions to improve this paper.

\bibliographystyle{alpha}
\bibliography{bibliography}

\newcommand{\etalchar}[1]{$^{#1}$}
\begin{thebibliography}{BMR{\etalchar{+}}99}

\bibitem[Baj88]{babaj88}
Chandrajit~L. Bajaj.
\newblock Geometric modeling with algebraic surfaces.
\newblock In {\em Proceedings of the 3rd {IMA} Conference on the Mathematics of
  Surfaces, Keble College, Oxford, September 19-21, 1988}, pages 3--48, 1988.

\bibitem[Bat01]{batsanov01}
S.S. Batsanov.
\newblock Van der {W}aals radii of elements.
\newblock {\em Inorganic Materials}, 37(9):871--885, 2001.

\bibitem[Bed92]{Bedi92}
Sanjeev Bedi.
\newblock Surface design using functional blending.
\newblock {\em Computer-Aided Design}, 24:505--511, 09 1992.

\bibitem[BGA04]{Barbier04}
Aur{\'e}lien Barbier, Eric Galin, and Samir Akkouche.
\newblock Complex skeletal implicit surfaces with levels of detail.
\newblock {\em Journal of WSCG}, 12(1-3), 2004.

\bibitem[Bli82]{blinn82}
James~F. Blinn.
\newblock A generalization of algebraic surface drawing.
\newblock In {\em SIGGRAPH '82: {P}roceedings of the 9th {A}nnual {C}onference
  on {C}omputer {G}raphics and {I}nteractive {T}echniques}, page 273, New York,
  NY, USA, 1982. ACM.

\bibitem[BMR{\etalchar{+}}99]{Bernardini99}
F.~Bernardini, J.~Mittleman, H.~Rushmeier, C.~Silva, and G.~Taubin.
\newblock The ball-pivoting algorithm for surface reconstruction.
\newblock {\em IEEE Transactions on Visualization and Computer Graphics},
  5(4):349--359, 1999.

\bibitem[Bon64]{bondi64}
A.~Bondi.
\newblock {V}an der {W}aals {V}olumes and {R}adii.
\newblock {\em J Phys Chem}, 68(3):441--451, 1964.

\bibitem[BX97]{babaj97}
Chandrajit~L. Bajaj and Guoliang Xu.
\newblock Spline approximations of real algebraic surfaces.
\newblock {\em Journal of Symbolic Computation}, 23(2-3):315--333, 1997.

\bibitem[CCD00]{chen}
F.L. Chen, C.S. Chen, and J.S. Deng.
\newblock Blending pipe surfaces with piecewise algebraic surfaces.
\newblock {\em Chinese Journal of Computers}, 23(9):911--916, 2000.

\bibitem[{\v{C}}S04]{Skala04}
Martin {\v{C}}erm{\'a}k and V{\'a}clav Skala.
\newblock Curvature dependent polygonization by the edge spinning.
\newblock In Antonio Lagan{\'a}, Marina~L. Gavrilova, Vipin Kumar, Youngsong
  Mun, C.~J.~Kenneth Tan, and Osvaldo Gervasi, editors, {\em Computational
  Science and Its Applications -- ICCSA 2004}, pages 325--334, Berlin,
  Heidelberg, 2004. Springer Berlin Heidelberg.

\bibitem[{\v{C}}S05]{Skala05}
Martin {\v{C}}erm{\'a}k and Vaclav Skala.
\newblock Polygonization of implicit surfaces with sharp features by
  edge-spinning.
\newblock {\em The Visual Computer}, 21:252--264, 05 2005.

\bibitem[{\v{C}}S07]{Skala07}
Martin {\v{C}}erm{\'a}k and Vaclav Skala.
\newblock Polygonisation of disjoint implicit surfaces by the adaptive edge
  spinning algorithm of implicit objects.
\newblock {\em Int. J. Comput. Sci. Eng.}, 3(1):45--52, July 2007.

\bibitem[DG11]{dias11}
S\'ergio~E.D. Dias and Abel~J.P. Gomes.
\newblock Graphics processing unit-based triangulations of blinn molecular
  surfaces.
\newblock {\em Concurrency and Computation: Practice and Experience},
  23(17):2280--2291, 2011.

\bibitem[DG15]{dias15}
S\'ergio~E.D. Dias and Abel~J.P. Gomes.
\newblock Triangulating molecular surfaces over a lan of gpu-enabled computers.
\newblock {\em Parallel Computing}, 42:35--47, February 2015.

\bibitem[DNJG17]{dias17}
S\'ergio~E.D. Dias, Quoc~Trong Nguyen, Joaquim Jorge, and Abel~J.P. Gomes.
\newblock Multi-gpu-based detection of protein cavities using critical points.
\newblock {\em Future Generation Computer Systems}, 67:430--440, February 2017.

\bibitem[Gom03]{2003-gomes}
Abel Gomes.
\newblock A concise b-rep data structure for stratified subanalytic objects.
\newblock In {\em Proceedings of the Eurographics/ACM SIGGRAPH Symposium on
  Geometry Processing \emph{(SGP'03), Aachen, Germany, June 23-25}}, pages
  83--93. Eurographics Association, 2003.

\bibitem[GRM99]{1999-gomes}
Abel Gomes, Chris Reade, and Alan Middleditch.
\newblock A mathematical model for boundary representations of n-dimensional
  geometric objects.
\newblock In {\em Proceedings of the 5th Symposium on Solid Modeling and
  Applications \emph{(SPM'99), Ann Arbor, Michigan, USA, June 8-11}}, pages
  270--277. ACM Press, 1999.

\bibitem[GVJ{\etalchar{+}}09]{Gomes09}
Abel Gomes, Irina Voiculescu, Joaquim Jorge, Brian Wyvill, and Callum
  Galbraith.
\newblock {\em Implicit Curves and Surfaces: Mathematics, Data Structures and
  Algorithms}.
\newblock Springer Publishing Company, Inc., 1st edition, 2009.

\bibitem[HH85]{Hoffmann85}
C.~Hoffmann and J.~Hopcroft.
\newblock Automatic surface generations in computer aided design.
\newblock {\em The Visual Computer}, 1(2):92--100, August 1985.

\bibitem[KBH06]{Kazhdan06}
Michael Kazhdan, Matthew Bolitho, and Hugues Hoppe.
\newblock Poisson surface reconstruction.
\newblock In {\em Proceedings of the 4th Eurographics Symposium on Geometry
  Processing \emph{(SGP'06), Cagliari, Sardinia, Italy, June 26-28}}, pages
  61--70, Aire-la-Ville, Switzerland, Switzerland, 2006. Eurographics
  Association.

\bibitem[LHRS05]{Levet05}
Florian Levet, Julien Hadim, Patrick Reuter, and Christophe Schlick.
\newblock Anisotropic sampling for differential point rendering of implicit
  surfaces.
\newblock In {\em WSCG The 13th International Conference in Central Europe on
  Computer Graphics, Visualization and Computer Vision}. UNION Agency - Science
  Press, Plzen, Czech Republic, 2005.

\bibitem[LPP06]{lazard}
S.~Lazard, L.~M. Pe{\~n}aranda, and S.~Petitjean.
\newblock Intersecting quadrics: an efficient and exact implementation.
\newblock {\em Computational Geometry: Theory and Applications},
  35(1-2):74--99, 2006.

\bibitem[Mur91]{Muraki91}
Shigeru Muraki.
\newblock Volumetric shape description of range data using ``blobby model''.
\newblock {\em SIGGRAPH Computer Graphics}, 25(4):227--235, July 1991.

\bibitem[NHK{\etalchar{+}}85]{Nishimura85}
Hiromitsu Nishimura, Masashi Hirai, Tsuyoshi Kawai, Tory Kawata, Isao
  Shirakawa, and Kengo Omura.
\newblock Object modelling by distribution function and a method of image
  generation.
\newblock {\em Transactions of the IEICE Japan}, J68-D(4):718--725, 1985.

\bibitem[PK89]{Pilz89}
Markus Pilz and Hussein Kamel.
\newblock Creation and boundary evaluation of {CSG}-models.
\newblock {\em Engineering with Computers}, 5(2):105--118, 1989.

\bibitem[RG06]{Raposo06}
Adriano~N. Raposo and Abel~J.P. Gomes.
\newblock Polygonization of multi-component non-manifold implicit surfaces
  through a symbolic-numerical continuation algorithm.
\newblock In {\em Proceedings of the 4th International Conference on Computer
  Graphics and Interactive Techniques in Australasia and Southeast Asia
  \emph{(GRAPHITE'06), Kuala Lumpur, Malaysia, November 29 - December 02}},
  pages 399--406, New York, NY, USA, 2006. ACM.

\bibitem[RG12]{Raposo12}
Adriano~N. Raposo and Abel J.~P. Gomes.
\newblock 3{D} molecular assembling of {B}-{D}{N}{A} sequences using
  nucleotides as building blocks.
\newblock {\em Graph. Models}, 74(4):244--254, July 2012.

\bibitem[RG14]{Raposo14}
Adriano~N. Raposo and Abel~J.P. Gomes.
\newblock {{E}fficient deformation algorithm for plasmid {D}{N}{A}
  simulations}.
\newblock {\em BMC Bioinformatics}, 15:301, Sep 2014.

\bibitem[RG15]{Raposo15}
Adriano~N. Raposo and Abel J.~P. Gomes.
\newblock is{D}{N}{A}: A tool for real-time visualization of plasmid {D}{N}{A}
  monte-carlo simulations in 3{D}.
\newblock In {\em Bioinformatics and Biomedical Engineering}, pages 566--577.
  Springer International Publishing, 2015.

\bibitem[RQG09]{raposo09}
Adriano~N. Raposo, J.~Queiroz, and Abel~J.P. Gomes.
\newblock Triangulation of molecular surfaces using an isosurface continuation
  algorithm.
\newblock In {\em Proceedings of the 2009 International Conference on
  Computational Science and Its Applications}, ICCSA '09, pages 145--153. IEEE
  Computer Society, 2009.

\bibitem[RV85]{Requicha85}
Aristides Requicha and Herbert Voelcker.
\newblock Boolean operations in solid modeling: Boundary evaluation and merging
  algorithm.
\newblock {\em Proceedings of the IEEE}, 73(1):30--43, 1985.

\bibitem[War89]{warren}
J.~Warren.
\newblock Blending algebraic surfaces.
\newblock {\em ACM Transactions on Graphics}, 8(4):263--278, 1989.

\bibitem[WMW86]{Wyvill86}
Geoff Wyvill, Craig McPheeters, and Brian Wyvill.
\newblock Data structure for soft objects.
\newblock {\em The Visual Computer}, 2(4):227--234, Aug 1986.

\bibitem[WZ00]{wu}
T.~R. Wu and Y.~S. Zhou.
\newblock On blending of several quadratic algebraic surfaces.
\newblock {\em Computer Aided Geometric Design}, 17:759--766, 2000.

\end{thebibliography}

\end{document}